# Concept of spinsonde for multi-cycle measurement of vertical wind profile of tropical cyclones

Chung-How Poh, and Chung-Kiak Poh

Persistence UAV Research, 23 Halaman York, 10450 Penang, Malaysia

Email: Chung-How@ieee.org

**Abstract**

Tropical cyclones and cyclogenesis are active areas of research. Chute-operated dropsondes jointly developed by NASA and NCAR are capable of acquiring high resolution vertical wind profile of tropical cyclones. This paper proposes a chute-free vertical retardation technique (termed as spinsonde) that can accurately measure vertical wind profile. Unlike the expendable dropsondes, the spinsonde allows multi-cycle measurement to be performed within a single flight. Proof of principle is demonstrated using the RealFlight® simulation software and results indicate that the GPS ground speed correlates with the wind speeds to within ±5 kmh$^{-1}$. This technique reduces flying weight and increases payload capacity by eliminating bulky chutes. Maximum cruising speed ($V_h$) achieved by the spinsonde UAV is 372 kmh$^{-1}$.

## 1. Introduction

A tropical cyclone forms over the warm ocean waters near the equator. It is a rotating storm consisting of organized system of clouds and thunderstorms and it requires a constant input of energy in the form of latent heat. Tropical cyclones often cause widespread damage when they make landfall due to high winds, torrential rainfall and storm surges [1–3]. In the vertical direction, winds are strongest near the surface and decay with height within the troposphere [4]. The super typhoon Haiyan that made landfall in Philippine in 2013 has sustained winds of 315 $kmh^{-1}$ with gusts as strong as 380 $kmh^{-1}$ [5]. Manned aircrafts have been used to fly into and around tropical cyclones at a relatively safe height of 700 mbar level (approximately 3 km) to make measurements of interior barometric pressure and wind speed [6,7]. Primary risk to manned flight is associated with downdrafts. A near-mishap occurred in 1989 when a Lockheed WP-3D Orion was flown into Hurricane Hugo [8].

Unmanned aerial vehicle (UAV) approach has been investigated to reduce risks to flight crews and to explore new capabilities. The Aerosonde has a wing span of about 3 m and it made its first operational flight back in 1995 [9]. It penetrated the eyewall of the Typhoon Longwang on a reconnaissance observation mission in 2005 and the in situ wind measurement was consistent with the Doppler weather radar [10]. The recent Aerosonde Mark 4.7 has a dash speed of 150 $kmh^{-1}$ at sea level [11]. NASA begun employing the Global Hawk UAVs in 2007 to study tropical cyclone using the Airborne Vertical Atmospheric Profiling System (AVAPS), or more commonly known as the dropsondes, which were developed by the U.S National Center for Atmospheric Research (NCAR) [12]. The NASA Global Hawk is ideal as it is capable of flight altitudes greater than 16764 m and flight duration of up to 30 hours [12]. The expendable dropsondes is designed to be dropped from an aircraft at altitude to measure tropical cyclone conditions as it falls to the surface. From 50000 ft (15240 m) down to sea level, the descent rate varied from 29.46 to 11.68 $ms^{-1}$ (5800 to 2300 $ftmin^{-1}$). The dropsondes were available in two versions, the AVAPS II sonde and the mini sonde. The mini sonde measured 30.5 cm in length and 4.7 cm in diameter, with a mass of 165 g. The dropsondes cost approximately USD 350 each and about 1000 to 1500 of them are used annually [13,14].

Given that the dropsondes require chutes and are expendable, we hereby propose an alternative concept known as the spinsonde. It can be viewed as an integration of an UAV and a sonde. It relies on the stall-spin maneuver to slow its vertical descent and no chute is needed. The concept is illustrated using a radio control (RC) model aircraft simulation software.

## 2. Materials and Methods

### 2.1. Concept of spinsonde

To measure the vertical wind profile, instead of deploying chutes as in the case of the dropsonde, the spinsonde UAV will execute an aerobatic maneuver known as the stall-spin to achieve retardation in the vertical descent rate. In other words, the spinsonde employs a chute-free vertical retardation technique. The main advantage is that the spinsonde UAV may exit the spin at any time, fly to another location and repeat the measurement sequence. In this way, it can acquire multiple vertical wind profile data in a single flight. Furthermore, eliminating the bulky chute will increase payload capacity.

### 2.2. Simulation details

Simulation work was performed using the RealFlight® 6.5 simulator [15] running on a quad-core 2.2 GHz computer. It employs RealPhysics™ technology for accurate and realistic simulations [15]. The as-supplied simulation model AR-6 Endeavor (Fig. 1) was used as the RC model to demonstrate the concept of the spinsonde. The single propeller-driven AR-6 Endeavor was chosen because it is Formula-One air racer and its high cruising speed will be an advantage in transversing tropical cyclones. The Endeavor model was powered by a 2-stroke 1.60 cu in glow engine, and it has a wing span of 2.23 m and flying weight of 4.96 kg. Wing chord at root and tip were 292 and 192 mm, respectively. Modification of the Endeavor was done using the Accu-Model™ aircraft editor. In order to further improve the airplane's better handling of the hurricane wind conditions, the wing span and chord were reduced so as to increase its wing-loading. It has a wing span of 1.94 m and the length of the horizontal stabilizer was shortened 32% to further reduce drag. The landing gear was eliminated as well. The modified version is codenamed SP-1. Note that during simulation run, the original graphical model of AR-6 will be displayed despite physical properties, such as wing span, have been altered. The graphical model will need to be updated separately but for this work it was not done as it does not affect the flight physics central to this study. Table 1 summarizes the key specifications of both aircrafts. The primary components added to the platform to enable electric-powered flight were a brushless motor and a 12-cell 16 Ah lithium polymer (Li-po) battery pack. The key advantages of employing brushless electric motor are reliability in the presence of rain water, and insensitivity to change in air pressure. 3-axis angular position dependent (roll, pitch, yaw) gyro was added to the platform for flight stabilization. The completed model weighed in at 6.73 kg with a wing loading of 368.6 g dm$^{-2}$. Flight performance and stability was evaluated under different environmental conditions.

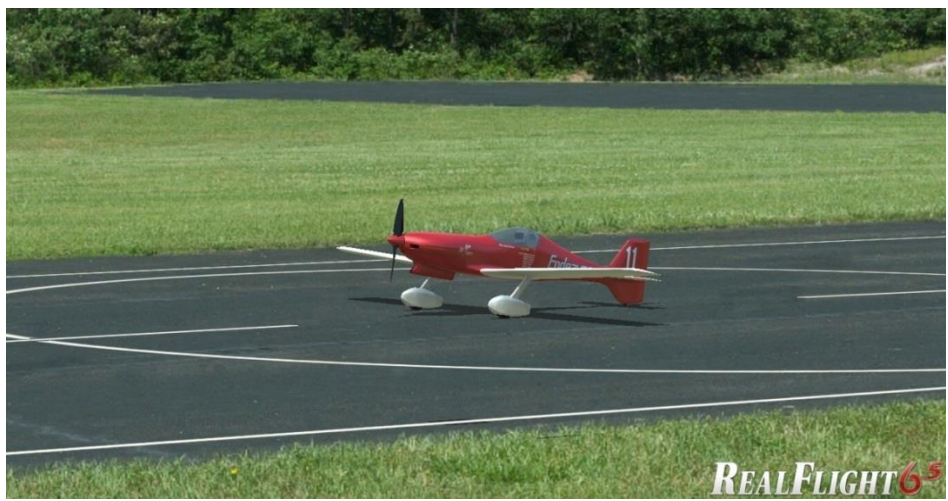

**Fig. 1** The AR-6 Endeavor model aircraft as supplied by the RealFlight® simulator.

**Table 1.** Specifications for the as-supplied AR-6 and the SP-1

| Aircraft model | Flying weight (kg) | Fuselage width (m) | Wing span (m) | Wing loading (g dm$^{-2}$) |
|---|---|---|---|---|
| AR-6 | 4.96 | 0.251 | 2.23 | 89.1 |
| SP-1 | 6.73 | 0.130 | 1.94 | 156.26 |

# 3. Simulation results and discussion

## *3.1. Level flight and handling characteristics*

The SP-1 achieved a $V_h$ (maximum speed in level flight with maximum continuous power) of 372 kmh$^{-1}$ which is faster than the sustained winds of the super Typhoon Haiyan (315 kmh$^{-1}$). This, in principle, would allow the SP-1 to be flown in tropical cyclones of similar strength to Typhoon Haiyan. The flight stability of the platforms under strong and gusty wind conditions was performed for different wind directions: headwind, tailwind, and crosswind. The platforms were subjected to horizontal wind speed of 225 kmh$^{-1}$, which is the maximum limit permitted by the simulator and with 100% turbulence and 30% wind gust. The control surfaces of the SP-1 were found to be responsive under these conditions with no adverse flight characteristics. Whenever there were perturbations to the roll, pitch and yaw angles, the onboard 3-axis heading hold gyro would attempt to steer them back. No inherent flight control instability was observed. Furthermore, the SP-1 was more resistant to turbulence than the AR-6 due to its higher wing loading.

## *3.2. Spinsonde and vertical wind profile measurement*

Chutes were used in the dropsondes to slow their vertical descent rate and the descent rate of the dropsondes close to sea-level was 11.68 ms$^{-1}$. This section investigates the effectiveness of spinsonde technique in measuring the wind profile. Eliminating the bulky chute will increase payload capacity and more importantly, enable the vertical descent sequence to be repeated multiple times per flight. The SP-1 was deliberately entered into a stall to initiate the maneuver. The steady-state stall-spin was achieved by applying full deflection of rudder and aileron in the same direction and applying up elevator. The angular displacements of all three control surfaces were approximately $45°$. The maneuver was performed under 500 m altitude and the descent rate was found to be 11.2 ms$^{-1}$, as indicated by the variometer as shown in Fig. 2. The stall-spin could be terminated by returning the control surfaces to neutral position and pulling up on the elevator to resume horizontal flight.

To investigate the ability of the technique to reliably detect wind speed, the stall-spin maneuver for the SP-1 was repeated with various horizontal wind speeds. Turbulence and gust were turned off in order to evaluate the effect reliably. One could observe the platform drifting in the lateral direction with the wind. Fig. 3(a) and (b) show the drifting in wind speed of 50 kmh$^{-1}$ and 180 km$^{-1}$, respectively. Ground speeds were found to closely match the wind speeds to within ±5 kmh$^{-1}$. This simulation work provided evidence that the stall-spin maneuver can be used as substitution for chute.

Videos showing the stall-spin maneuvers of the SP-1 including the exit to level flight in still air and strong wind are available as ancillary files ([1](#) and [2](#), respectively). Also, it can be observed from the movie files that the axis of rotation of the stall-spin is close to the center of gravity of the aircraft and, therefore the maneuver is not expected to affect the empirical GPS speed measurement as the spatial variation is smaller than the resolution of GPS.

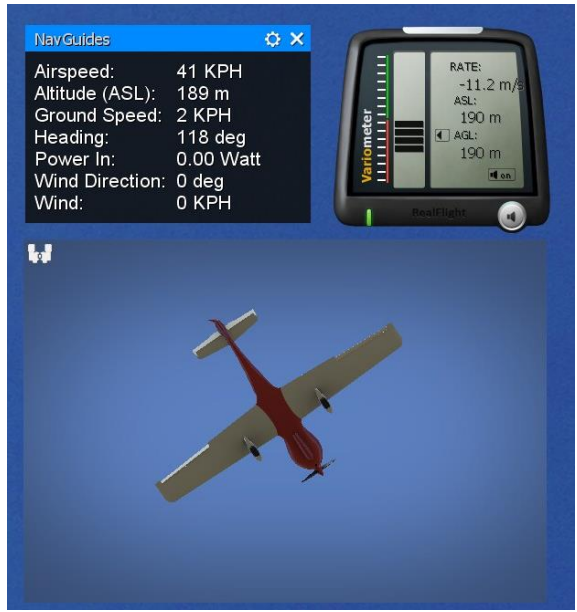

**Fig. 2** Dynamic parachuting: stall-spin maneuver performed by the SP-1 with the variometer showing a descent rate of 11.2 ms$^{-1}$. Full deflection of the control surfaces can be observed.

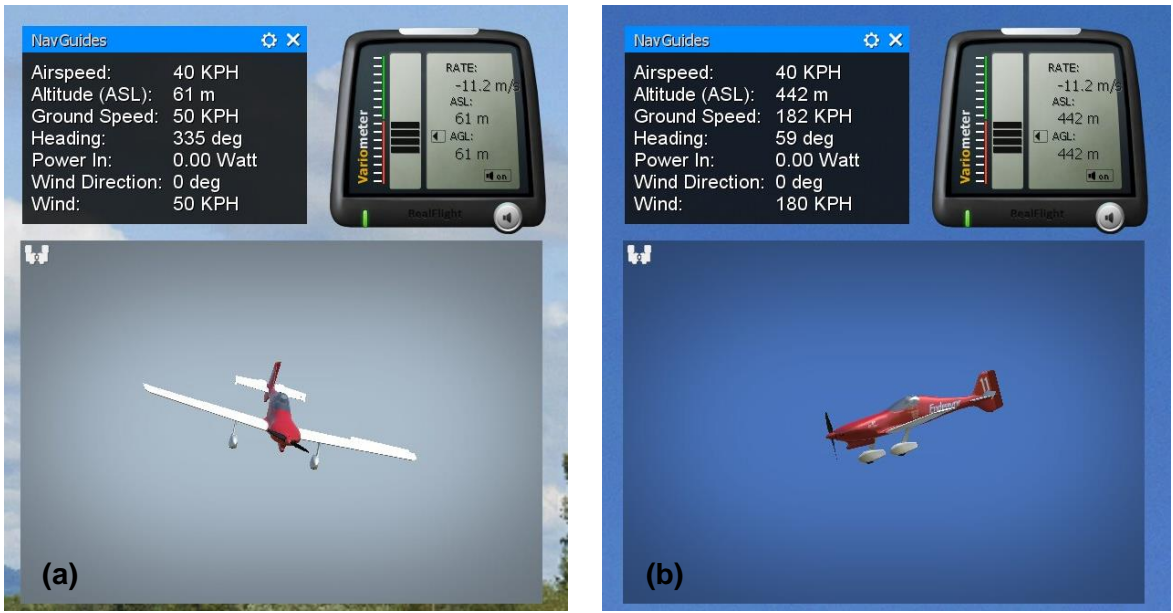

**Fig. 3** SP-1 performing the stall-spin maneuver amidst wind of intensity a) 50 kmh$^{-1}$ and b) 180 kmh$^{-1}$. The ground speeds were found to correlate strongly with the wind intensities to within ±5 kmh$^{-1}$.

## Conclusions

This work proposed and demonstrated via simulation the concept of spinsonde. It represents a chute-free technique applicable to UAV and is capable of performing vertical wind profile measurement within and around tropical cyclones. In still air, the simulated spinsonde UAV achieved a maximum cruising speed ($V_h$) of 372 kmh$^{-1}$. The proposed stall-spin maneuver was found to be an effective substitution for chute with the ability to carry out multi-cycle measurement at various target locations compared to the current expendable dropsonde method. Elimination of the chute also increases payload mass and volume. We believe the technique of spinsonde will open a new chapter in tropical cyclone research.